\documentclass{svmult}

\usepackage{graphicx}    

\def\beq{\begin{equation}}
\def\eeq{\end{equation}}

\def\bea{\begin{eqnarray}}
\def\eea{\end{eqnarray}}
\def\ba{\begin{array}}                  
\def\ea{\end{array}}

\begin{document}

\title*{Charge and Isospin Fluctuations in High Energy pp-Collisions}
\author{Mladen Martinis\inst{1}\and
Vesna Mikuta-Martinis\inst{2}}
\institute{Rudjer Bo\v skovi\' c Institute, Zagreb, Croatia
\texttt{martinis@rudjer.irb.hr} \and Rudjer Bo\v skovi\' c
Institute, Zagreb, Croatia \texttt{vmikuta@rudjer.irb.hr}}
%
%
\maketitle

Charge and isospin  event-by-event fluctuations in high-energy
pp-collisions are predicted within the Unitary Eikonal Model, in
particular the fluctuation patterns of the ratios of
charged-to-charged and neutral-to-charged pions. These
fluctuations are found to be sensitive to the presence of unstable
resonances, such as  $\rho $ and $\omega $ mesons.  We predict
that the charge-fluctuation observable $D_{UEM}$ should be
restricted to the interval $8/3\le D_{UEM}\le 4$ depending on the
$\rho /\pi $ production ratio.  Also, the isospin fluctuations of
the DCC-type  of the ratio of neutral-to-charged pions are
suppressed if pions are produced together with $\rho $ mesons.

\section{Introduction}
\label{sec:1}

In a single central ultrarelativistic collisions at RHIC and LHC
more then 2400 hadrons are created \cite{Sikler}, presenting
remarkable opportunity to study event-by-event fluctuations of
various hadronic observables.  Such  single event analysis with
large statistics may reveal new physical phenomena usually hidden
when averages over a large statistical sample of events are made
\cite{hei}. Recently, the study of event-by-event fluctuations of
charged particles in high-energy pp and heavy-ion collisions has
gained a considerable attention \cite{jk,ahm,hj,gpz,rev}. The idea
was to find an adequate  measure that can differentiate a
quark-gluon plasma (QGP) from a hadron gas (HG). So far, no
consideration  has been given to the fluctuations generated by the
phase transition (PT) itself \cite{hwa1}.

The number of particles produced in relativistic pp and heavy-ion
collisions can differ dramatically from collision to collision due
to the variation of impact parameter (centrality dependence),
energy deposition (leading particle effect), and other dynamical
effects \cite{3}. The fluctuations can also be influenced by novel
phenomena such as the formation of disoriented chiral condensates
(DCCs) \cite{4,5,16,17,18,19,20,23} in consequence of the
transient restoration of chiral symmetry. It is generally accepted
that much larger fluctuations  of the  neutral-to-charged pion
ratio than expected from Poisson-statistics
 could be a sign of the DCC formation.
However, such fluctuations  are possible even without invoking the
DCC formation if, for example, pions are produced semiclassically
and constrained by global conservation of isospin
\cite{10,11,12,13,14}. In this paper, we present results of  an
event-by-event  analysis of charged-charged and neutral-charged
pion fluctuations  as a function of the $\rho /\pi $ production
ratio in pp-collisions. Our study of these fluctuations is
performed within the Unitary Eikonal Model (UEM) \cite{24,34}.

\section{Coherent production of $\pi $ and $\rho $-mesons}
\label{sec:2}

At high energies most of the pions are
produced in the nearly baryon-free central region of the phase space.
The energy available for their production is
\begin{equation}
E_{had} = \frac{1}{2} \sqrt{s} - E_{leading}
\end{equation}
which at fixed total c.m. energy $ \sqrt{s} $ varies from
event-to-event. Within the UEM the N-pion contribution to the
$s$-channel unitarity in the central region can be written as an
integral over the relative impact parameter  $\vec{b}$ between two
incident leading particles:
\begin{equation}
 \sigma_{N}(s) = \frac{1}{4s^2} \int d^{2}b \prod_{i=1}^{N}dq_{i}
\mid T_{N}(s, \vec{b};q_1 \ldots q_N) \mid^{2},
\end{equation}
where $ dq = d^{2}q_{T}dy/2(2 \pi)^{3}.$ If the isospin  of the
incoming (outgoing) leading particle-system  is $II_{3}$
($I'I'_{3}$), then the $N$-pion production amplitude becomes
cite{24}
\begin{equation}
iT_{N}(s, \vec{b};q_{1} \ldots q_{N}) = 2s \langle I'I'_{3};q_{1}
\ldots q_{N} \mid \hat{S}(s, \vec{b}) \mid II_{3} \rangle,
\end{equation}
where $\hat{S}(s, \vec{b})$ denotes the $\hat{S}$- matrix in the
isospace of the leading particles.

The coherent emission of  pions or clusters of pions, such as
$\rho $ and $\omega $ mesons, in b-space of leading particles is
described by the factorized form of the scattering amplitude,
$T_{N}$. In that case the $ \hat{S}(s, \vec{b})$-matrix has the
following generic form:
\begin{equation}
\hat{S}(s, \vec{b}) = \int d^{2} \vec{n} \mid \vec{n}
\rangle \hat{D}( s, \vec{b}) \langle \vec{n} \mid,
\end{equation}
where $ \mid \vec{n} \, \rangle $ represents the isospin-state
vector of the two-leading-particle system.
The quantity $\hat{D}( s, \vec{b})$ is the unitary coherent-state
displacement operator defined in our case as
\begin{equation}
D( s, \vec{b}) = exp  [a^{\dagger}(s, \vec{b}) - a(s, \vec{b})]
\end{equation}
with
\begin{equation}
 a^{\dagger}(s, \vec{b}) =  \sum_{c=\pi ,\rho }\int dq
 J_{c}(s, \vec{b};q)  \vec{n}  \vec{a_{c}}^{ \dagger}(q).
\end{equation}
where, $ J_{c}$ denotes  a classical source function of the
cluster $c$. The cluster decays into  pions outside the region of
strong interactions (i.e. the final-state interaction between
pions is neglected).

The isospin $(I', I_{3}')$ of the outgoing leading particle system
varies from event-to-event. If  the  probabilities $\omega_{I',
I_{3}'}$ of producing $(I', I_{3}')$ states are known, we can sum
over all $(I'I'_{3})$
 to obtain a  probability distribution of producing $N_{+}, \,
N_{ \_} $ and $N_{0}$ pions from a given initial isospin state:
\begin{eqnarray}
P_{I I_{3}}(N_{+}N_{ \_}N_{0}\mid N)  C_{I I_{3}}(N) &  =  & \nonumber \\
& & \hspace*{-4cm} \sum_{I'I'_{3}}\omega_{I', I_{3}'} \int d^{2}b dq_{1}
dq_{2} \ldots dq_{N} \mid \langle I'I'_{3},N_{+}N_{ \_}N_{0} \mid
\hat{S}(s,
\vec{b}) \mid II_{3} \rangle \mid^{2})
\end{eqnarray}
where
\begin{eqnarray}
 N &  =  & N_{+} + N_{ \_} + N_{0}\nonumber \\
 N_{+} &  = & n_{\pi ^+} + n_{\rho ^+} + n_{\rho ^0} \nonumber\\
 N_{-} &  = & n_{\pi ^-} + n_{\rho ^-} + n_{\rho ^0} \nonumber \\
 N_{0} &  = & n_{\pi ^0} + n_{\rho ^+} + n_{\rho ^-}
\end{eqnarray}
and $ C_{I I_{3}}(N)$ is the corresponding normalization factor.
This is now our basic equation for calculating various
pion-multiplicity distributions, pion-multiplicities, and
pion-correlations between definite charge combinations. In the
following, we consider fluctuations of the $\pi_{+}/\pi_{-}$ and
$\pi_{0}/N$ ratios in pp-collisions, $(I=I_{3}=1)$.
\section{Charge and isospin fluctuations}
\label{sec:3}

A suitable measure of the charge fluctuations was suggested in
\cite{jk}. It is related to the fluctuation of the ratio $R_{ch} =
N_{+}/N_{-}$ and the observable to be studied is
\begin{equation}
D \equiv <N_{\rm ch}><\delta R^2_{\rm ch}> =
 4 \frac{<\delta Q^2> }{ <N_{\rm ch}>}.
\end{equation}
where $N_{\rm ch} = N_{+} + N_{-},\, Q = N_{+} - N_{-}$ and
$<\delta Q^2> = <Q^2> - <Q>^2$.

Our prediction of the D quantity within the UEM, when both $\pi $
and $\rho $ mesons are produced is

\begin{equation}
D_{UEM} = \frac{8}{2 + \frac{\overline{n}_{ \pi}}{\overline{n}_{
\pi} + \overline{n}_{ \rho}}} ,
\end{equation}

where $ \overline{n}_{ \pi} = \int dq \mid J_{\pi }(q)\mid^2$
denotes the average number of  directly produced pions, and
similarly $\overline{n}_{ \rho}$ denotes the average number of
$\rho$ mesons which decay into two short-range correlated pions.
The total number of emitted pions is
\begin{equation}
\overline N = \overline{n}_{ \pi} + 2 \overline{n}_{ \rho}.
\end{equation}

It was argued \cite{jk} that the value of D may be used to
distinguish the hadron gas ($D_{\rm \pi-gas} \approx 4 $) from the
quark-gluon plasma ($D_{\rm QGP} \simeq 3/4 $).It is expected that
$D_{\rm \pi-gas} \simeq 3$ if appropriate corrections for
resonance production are taken into account \cite{koch}.

The UEM predicts $D_{UEM} = 4$ if $\overline{n}_{ \rho} = 0$. In
that case  the   pion production  is restricted only by the global
conservation of isospin. However, if $\overline{n}_{ \pi} = 0$ the
UEM predicts $D_{UEM} = 8/3$.This means that D is restricted to
the interva $8/3\le D_{UEM}\le 4$.

The preliminary results from CERES , NA49  and STAR collaboration
\cite{ce,49,st}, however,  indicate that the measured value of $D$
is close to that predicted for hadron gas and differs noticeably
from that expected for QGP. This finding is somewhat disturbing
since no effect of resonance production is visible in the
fluctuations.

The  formation of DCC in pp-collisions is expected to lead to
different types of isospin fluctuations \cite{16,17,18,19,20}.
Since  pions formed in the DCC are essentially classical and form
a quantum superposition of coherent states with different
orientation in isospin space,large event-by-event fluctuations in
the ratio  $R_{0} = N_{0}/N$ are expected.  The probability
distribution of $R_{0}$ inside  the DCC domain is
\cite{13,16,17,19}
\begin{equation}
P_{DCC}(R_{0}) = \frac{1}{2\sqrt{R_{0}}}
\end{equation}
 There are a variety of proposed mechanisms
other than DCC formation which can also lead to the distribution
$P_{DCC}(R_{0})$ \cite{23,24,25,26}. The distribution $
P_{DCC}(R_{0})$ is different from the generic
binomial-distribution expected in normal events which assumes
equal probability for production of $\pi _{+}$, $\pi _{-}$ and
$\pi _{0}$ pions.

Following the approach of our earlier papers \cite{24,34}, we have
calculated the probability distribution function,
 $P^{\pi + \rho}(N_{0}\mid N)$ for producing $N_{0}$  neutral pions.
For large $N$ and $N_{0}$ such that $R_{0}$ is fixed,we find that
only $NP^{\pi}(N_{0}\mid N)$ is of the form $\sqrt{N / N_{0}}$
which resembles the DCC-type fluctuations and is typical for
coherent pion production.

\section{Conclusion}

Our general conclusion is that within the UEM the large charge
and  isospin fluctuations depend strongly on the value of the
$\rho /\pi $ production ratio which fluctuate from event to event.
Recent estimate of the $\rho /\pi $ production ratio at
accelerator energies, is $ \overline{n}_{ \rho} = 0.10
 \overline{n}_{ \pi}$ \cite{jk}.

\end{document}